# STATUS OF THE CONTROL SYTEM FOR THE FRONT-END OF THE SPALLATION NEUTRON SOURCE {*}

S. A. Lewis, C. A. Lionberger,
P. T. Cull, LBNL, Berkeley, CA 94720, USA

Abstract

The Spallation Neutron Source (SNS) is a partnership between six laboratories. To ensure a truly integrated control system, many standards have been agreed upon, including the use of EPICS as the basic toolkit. However, unique within the partnership is the requirement for Lawrence Berkeley National Lab, responsible for constructing the Front End, to operate it locally before shipping it to the Oak Ridge National Lab (ORNL) site. Thus, its control system must be finished in 2001, well before the SNS completion date of 2006. Consequently many decisions regarding interface hardware, operator screen layout, equipment types, and so forth had to be made before the other five partners had completed their designs. In some cases the Front-End has defined a standard by default; in others an upgrade to a new standard is anticipated by ORNL later. Nearly all Front-End devices have been commissioned with the EPICS control system. Of the approximately 1500 signals required, about 60% are now under daily operational use. The control system is based on "standard architecture"; however, it has a field-bus dominated layout. This paper will discuss some unique interface requirements that led to adding new device families into the EPICS repertoire. It will also describe the choices and trade-offs made for all major areas.

## 1 OVERVIEW

The Spallation Neutron Source (SNS) is a 1 MW, pulsed neutron source being built by six Department of Energy Labs and scheduled for completion at Oak Ridge National Lab (ORNL) in 2006. At completion it will cost about $1.4B and extend over nearly 1 km. Lawrence Berkeley National Laboratory (LBNL) is tasked with constructing the Front End (FE) at Berkeley by June, 2002 and then shipping it to ORNL for re-commissioning. The FE itself is comprised of three sections: the Source and Low Energy Beam Transport (LEBT) which produces 50 mA of 65 keV H$^-$ suitably chopped and focussed (using electrostatic devices); the Radio Frequency Quadrupole (RFQ) which bunches the beam with 800 kW of 402 MHz RF and further accelerates it to 2.5 MeV; and the Medium Energy Beam Transport, which matches the beam to the Linac (using a magnetic lattice) and includes diagnostic devices. All three sections are served by suitable vacuum pumping, employing turbo-molecular, cryogenic, and getter-ion types.

Thus it can be seen that although the FE represents roughly 3% of the full SNS measured by beam energy (2.5 MeV), length (10 m), or cost ($20M), it has nearly the same complexity. For the FE control system, a similar ratio applies: it will have about 1550 signals (of about 60,000 for SNS), cost $1.5M (of about $60M); have one operator console (of 12), and three I/O controllers (of 150).

When complete, the 1550 signals in the FE control system will come from 300 devices, assigned by system: Source/LEBT, 250; RFQ, 100; MEBT, 700; Vacuum, 500.

To support the extensive R&D program and operation of the whole FE, the control system has been in continuous operation since 1999, with incremental additions to track the FE itself. As of 1 November, two of the three I/O controllers (IOCs) are in service, with about 700 signals (for 200 devices) interfaced, essentially completing the Source, LEBT, and RFQ systems. In addition to the required operator console (OPI), three temporary OPIs have been provided to allow concurrent but independent operation of the 402 MHz systems, vacuum commissioning, Source/LEBT testing, and a software development area. An EPICS "gateway" allows controls and management staff at SNS partner labs to continuously and efficiently monitor active devices, while critical engineering and technical staff routinely monitor and control devices in their jurisdiction from their offices or even homes.

## 2 FRONT-END AND SNS STANDARDS

The SNS has chosen the well-known EPICS toolkit [1] as the basis for its control system. While assuring easy integration among the multiple collaborators, this decision still leaves open many options. Because of the early start and immediate operational needs of the FE, it did not adhere to all of the final SNS standards as developed by the collaborating partners [2].

---

Work Supported by the U.S. Department of Energy under Contract #DE-AC05-00OR22725

## 2.1 Hardware

**IOC**. The EPICS Input/Output Controller (IOC) uses 21-slot Wiener™ or 7-slot Dawn™ VME64x or VXI crates with Motorola™ MVME-2100 series PowerPC CPUs. FE will conform to the SNS standard (although it has operated well to date with older MVME-167 68K series). SNS uses the VxWorks™ kernel.

**PLC**. For robust vacuum service, a layer of Programmable Logic Controller (PLC) is placed between the EPICS IOCs and the hardware interface. FE has chosen the Allen-Bradley™ (A-B) PLC/5 family, with IOC-to-PLC and PLC-to-I/O both using the proprietary A-B Remote-I/O (RIO; RS-232 based) and the 6008 VME Scanner; and the A-B 1794 (Flex-I/O™) interface modules. FE will not conform to the SNS standard[3], which stays with the A-B family, but uses the ControlLogix/5000™ family; Ethernet/IP for IOC-to-PLC and ControlNet™ for PLC-to-I/O, as well as CLX/5000 family interface modules.

FE exclusively uses self-contained vacuum gauge units. FE does not conform to SNS standards, which require rack-mounted electronics away from high radiation areas.

**Power Supply I/O**. For general electrostatic, magnetic, RF and other devices, FE has two alternate specifications (see ¶4): (1) Group-3, a proprietary system utilizing a serial fiber ring; (2) A-B Flex-I/O directly driven from A-B 6008 Scanner with RIO. FE will not conform with the SNS standard[4], which uses a BNL-designed Power Supply Controller/Interface technology.

**Motion Control**. FE will use OMS-58 VME boards. FE will conform to SNS standards.

**High-Power RF**. The temporary interface now in use will be replaced by a LANL-supplied, VXI-based unit including full EPICS support, which is the SNS standard.

## 2.2 Software

**Application Development Environment (ADE)**. The ADE is an enhancement of the EPICS application *makefiles* with a strong notion of release management (based heavily on experience at TJNAF with the CEBAF EPICS system) that recognizes the common usage of shared drivers across applications, and of shared applications among many IOCs. All textual file entities are version-controlled using CVS at a single repository at the SNS ORNL site. FE conforms to the SNS standard.

**Operator Screens (HMI).** EPICS offers several "display manager" client applications. FE conforms to the initial SNS standard, *dm2k*. However, a recent change to *edm* will require an upgrade. FE conforms to use of all other EPICS HMI-level clients such as *Knob Manager*, *StripTool*, *Channel Archiver*, *Backup Restore Tool*, *Probe*, *etc.* FE initiated a suite of colors, fonts, navigation strategy, layout and other HMI visual factors which have become the SNS standard.

**Sequencing**. The EPICS sequencer (v 1.9.5) is used for all applications which are well modelled as a Finite State Machine. In particular, it is used for high-powered RF conditioning; for RFQ temperature-frequency controls; for remote oscilloscope controls; and is under development for vacuum sequencing. FE conforms to the SNS standard.

**Naming**. A naming system was adopted by a working group for controls well before FE construction started. FE conforms to the SNS standard for systems, device, and signals (some late revisions will be addressed as an upgrade).

**Relational Database**. FE uses structured text files with extensive macro-substitution scripts to expand to the full EPICS configuration from primary files. FE does not conform to the SNS standard Oracle relational database to generate the configuration nor to hold the primary data in tables. Until an upgrade is performed, no mechanism exists to ensure compliance with the Naming standard.

## 3 FIELD-BUS DOMINATED LAYOUT

The FE control system at the level of block diagram is a conventional, "standard architecture" as used in both general and EPICS contexts; however, as the choice of either Group-3 or Flex-I/O for all device interfaces became pervasive, it became clear that overall this would be a field-bus dominated layout. That is, no analog or digital signals are directly wired to or from the VME modules; (except OMS-58 motor controller cables); rather, all device signals are distributed over robust, serial, "daisy-chain" media via a protocol.

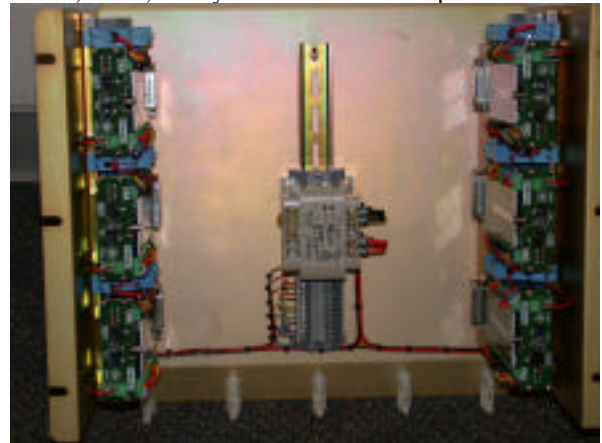

Fig 1: Rear rack U-chassis for Group-3.

Thus, there are no cross-connects; no requirement for low-level cable isolation in wire-ways; and no congestion as 700-800 signals converge in the racks housing the VME crates. Instead, outboard units were fabricated, either as rack-drawers immediately adjacent to rack-mounted equipment; as U-shaped chassis mounted on the rear rack rails (Fig 1); or as enclosed beam-line mounted boxes (Fig 2). All three types were heavily re-used, requiring only different wire-lists; and they all are fully connectorized to allow easy servicing.

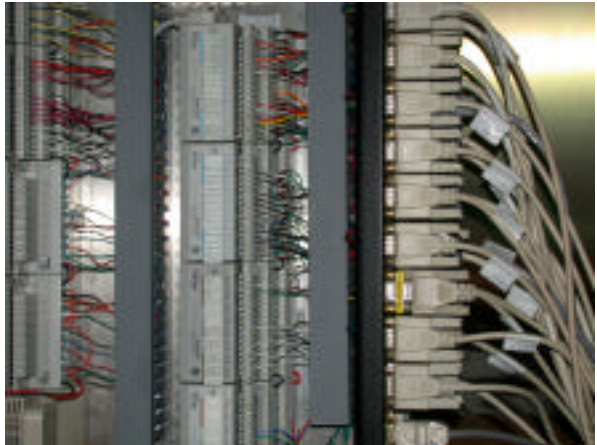

Fig 2: Enclosure for Flex-I/O with connectors.

## 4 NEW DEVICE TYPES FOR EPICS

### 4.1 Group-3

The Group-3 technology was chosen for its easy isolation of high potentials found in the Source and LEBT areas as well as for noise immunity from the severe transients expected due to sparking. The isolation is a function of the fiber which connects the individual units on the proprietary serial field bus; the noise immunity is improved by the liberal inclusion of typical bypass circuitry. As advised by the vendor, all due care is still required in external circuitry, grounding, and shielding practices.

The Group-3 technology is also used where the specific interface modules are particularly well matched to the device. The CNA unit has a combination of high-resolution analog with filtering combined with digital signal handling which gives a one-to-one match with most LEBT and MEBT DC power supplies. The more configurable CN3 unit is used for motor/encoder requirements or cases with unusual input/output combinations. Driver software for this new EPICS device family was added by FE during the R&D phase, and includes very detailed, online, concurrent device diagnostics (Fig 3).

### 4.2 Allen-Bradley Flex-I/O

Flex-I/O is a moderate-cost, lower-resolution option for industrial type signals (RFQ and MEBT cooling) that can share the A-B 6008 scanner and RIO cabling with the PLC. Although EPICS has long had a rich repertoire of drivers for it, more detailed concurrent online diagnostics were added based on the favorable experience with Group-3, and drivers for several newer 1794 family items were added.

### 4.3 Tektronix™ Scopes

Both low- and high-level support was provided for the Tektronix TDS200 and TDS3000 2- and 4-trace scope families, using RS-232 and a Janz™ 8-port serial VME card for the former, and Ethernet for the latter. A near replica of the scope front panels is available to the operator. The serial line units update at about 1 trace/sec, whereas the Ethernet models can achieve about 25 traces/sec (500-points/trace).

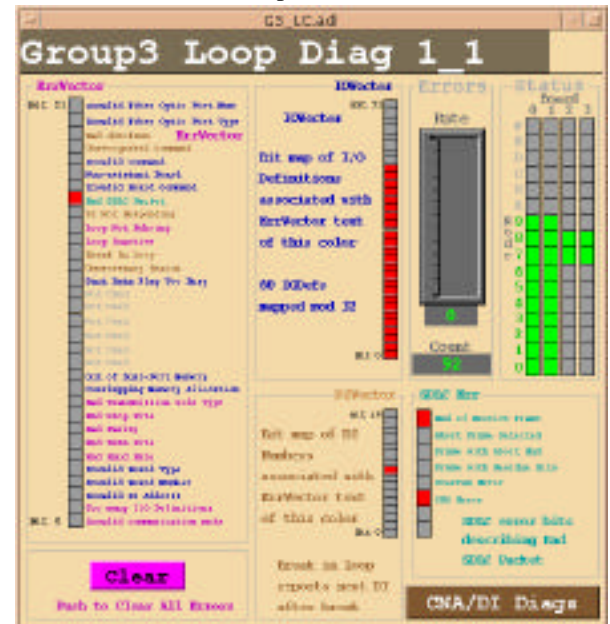

Fig 3: Diagnostics screen for Group-3 devices.